# Head/tail Breaks for Visualization of City Structure and Dynamics

Bin Jiang

Faculty of Engineering and Sustainable Development, Division of Geomatics
University of Gävle, SE-801 76 Gävle, Sweden
Email: bin.jiang@hig.se

**Abstract**
The things surrounding us vary dramatically, which implies that there are far more small things than large ones, e.g., far more small cities than large ones in the world. This dramatic variation is often referred to as fractal or scaling. To better reveal the fractal or scaling structure, a new classification scheme, namely head/tail breaks, has been developed to recursively derive different classes or hierarchical levels. The head/tail breaks works as such: divide things into a few large ones in the head (those above the average) and many small ones (those below the average) in the tail, and recursively continue the dividing process for the large ones (or the head) until the notion of far more small things than large ones has been violated. This paper attempts to argue that head/tail breaks can be a powerful visualization tool for illustrating structure and dynamics of natural cities. Natural cities refer to naturally or objectively defined human settlements based on a meaningful cutoff averaged from a massive amount of units extracted from geographic information. To illustrate the effectiveness of head/tail breaks in visualization, I have developed several case studies applied to natural cities derived from the points of interest, social media location data, and time series nighttime images. I further elaborate on head/tail breaks related to fractals, beauty, and big data.

## 1. Introduction

The things surrounding us vary dramatically, which implies that, instead of more or less similar things, there are actually far more small things than large ones, e.g., far more small cities than large ones in the world. This dramatic variation is often referred to fractal or scaling, and is well captured by geographic information of various kinds. Reflected in the points of interest (POI), there are far more POI in cities than in countryside; in terms of social media, there are far more users in cities than in the countryside; seen from the time-series nighttime images, there are far more dark/gray pixels than light ones. The new kind of geographic information constitutes what we now call big data (Mayer-Schonberger and Cukier 2013) in contrast to conventional small data. Unlike small data (e.g., census or statistical data), which are often estimated and aggregated, geographic information in the big data era is accurately and precisely measured at an individual level. This kind of geographic information due to its diversity and heterogeneity is likely to show the scaling pattern of far more small things than large ones. To better reveal the scaling structure, a new classification scheme, namely head/tail breaks (Jiang 2013), has been developed to recursively derive inherent classes or hierarchical levels. It divides things around an average, according to their geometric, topological and/or semantic properties, into a few large ones in the head (those above the average) and many small ones (those below the average) in the tail, and recursively continues the dividing process for the large ones (or the head) until the notion of far more small things than large ones has been violated (c.f., Section 2 for a working example).

Natural cities refer to naturally and automatically derived human settlements, or human activities in general on the earth's surface, based on a meaningful cutoff averaged from a massive amount of units extracted from massive geographic information. For example, we build up a huge triangulated irregular network (TIN) consisting of one-day tweets locations indicated by GPS coordinates rather place names around the world. It is obvious that with the TIN there are far more short edges than long ones. The average length of the edges splits all the edges into two parts: a minority of long edges (longer than the average) in the head, and a majority of short edges (shorter than the average) in the tail of the rank-size plot (Zipf 1949). Aggregate all short edges to create thousands of natural cities around the world. The natural cities are a collective decision of diverse, independent, and heterogeneous TIN edges, thus manifesting some wisdom of crowds (Surowiecki 2004). Interestingly, the natural cities demonstrate striking fractal structure and nonlinear dynamics (Jiang and Miao 2014). While conventional cities imposed by authorities from the top down are of great use for administation and mangement, natural cities defined from the bottom up are of more use for studying the underlying structure and dynamics. Natural cities are not constrained to individual countries, but are universally defined and delineated for the entire world with support of big data. Because of the universality, natural cities defined at very fine spatial and temporal scales provide a useful means for scientific research.

This paper attempts to develop an argument that in the big data era head/tail breaks can become an efficient and effective visualization tool for illustrating structure and dynamics of natural cities. The fundamental logic of this argument is as such. A large number of natural cities as a whole can be classified into different hierarchical levels or classes. Instead of showing all the classes or the whole, we can deliberately drop out some low classes, yet without distorting the underlying scaling pattern of the whole. This is because the remaining classes as a sub-whole are self-similar to the whole. This logic applies to the time dimension as well, i.e., instead of showing all evolving patterns along a time line, we deliberately choose a part that reflects the whole. Head/tail breaks provides a simple instrument that helps us see fractals in nature and society, i.e., through examining whether there are far more small things than large ones, or more precisely whether the scaling pattern recurs multiple times. Conventionally, we must compute the fractal dimension to determine whether a set or pattern is fractal (Mandelbrot 1982). Fractal dimension (D) is rigorously defined, referring to the ratio of the change of details (N) to that of measuring scale (r), $D = \log(N)/\log(r)$. Following the rigorous definition, fractals are found to appear in a variety of phenomena such as mountains, trees, clouds, rivers, cities, streets, architectures, the Internet, the World Wide Web, social media, and even the paintings of Jackson Pollock (e.g., Batty and Longley 1994, Eglash 1999, Taylor 2006). Now we can simply judge the ubiquity of fractals relying on our intuitions, i.e., a set or pattern is fractal if the scaling pattern of far more small things than large ones recurs multiple times.

The remainder of this paper is structured as follows. Section 2 introduces head/tail breaks and discusses how it leads to a new definition of fractals using the Sierpinski carpet and Mandelbrot set as working examples. Section 3 reports several case studies applied to visualization of natural cities derived from POI, social media data, and nighttime images. Section 4 adds some further discussions on head/tail breaks to meet challenges from big data. Finally Section 5 concludes the paper, and points to future work.

**2. Head/tail breaks leading to a new definition of fractals**
Head/tail breaks is largely motivated by heavy-tailed distributions such as power law, lognormal, and exponential distributions (c.f. Section 4 for a discussion) to derive inherent classes or hierarchical levels. The resulting number of classes is given by another term called ht-index (Jiang and Yin 2014), as an alternative index to fractal dimension for characterizing the complexity of fractals. The higher the ht-index, the more complex the fractals. Before illustrating its visualization capability, I shall briefly introduce head/tail breaks using the working example of the Sierpinski carpet.

The Sierpinski carpet, as a classic plane fractal, contains far more small squares than large ones, i.e., 1, 8, and 64 squares with respect to sizes 1/3, 1/9 and 1/27 given the carpet of one unit (Figure 1). The fractal dimension of the Sierpinski carpet can be calculated by $D = \log(8)/\log(3) = 1.893$, which indicates that every time the scale (r) is reduced three times, the number of squares (N) increases eight times. The calculation may look somewhat abstract and hard to grasp. Now let us take a simpler and easier way. There are far more small squares than large ones; at the smallest end there are 64 squares sized 1/27, at the largest end 1 square sized 1/3, and in the middle of the two ends 8 squares sized 1/9. If we create a scatterplot of these three points in an Excel sheet and fit them into a power function, one would observe y = 0.125 x ^ -1.893 (see Figure 1). This is called the Richardson plot, showing the ratio of the change of details (N) to the change of scales (r). In the Richardson plot, three points are exactly on the distribution line, implying that the Sierpinski carpet is strict fractal, or alternatively, the parts are strictly self-similar to the whole. If we replaced the squares with city sizes, the points would be around rather than exactly on the distribution line. This is because city sizes are just statistically fractal rather than strictly fractal.

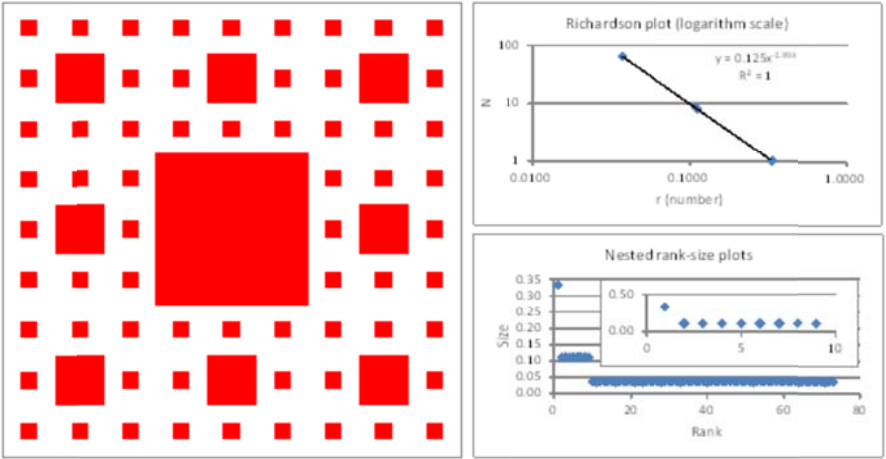

Figure 1: (Color online) Illustrstion of head/tail breaks and fractal dimension using the Sierpinski carpet
(Note: There are far more small squares than large ones for the Sierpinski carpet. The Richardson plot shows the fractal dimension, while the nested rank-size plots demonstrate the head/tail breaks process or the ht-index

Table 1: The head/tail breaks for the Sierpinski squares

| # Squares | Mean | # head | # tail | % head | % tail |
|---|---|---|---|---|---|
| 73 | 0.0492 | 9 | 64 | 12% | 88% |
| 9 | 0.1358 | 1 | 8 | 11% | 89% |

Now let us examine how head/tail breaks works for the Sierpinski carpet. There are a total of 1 + 8 + 64 = 73 squares, and the average size of which is calculated by $m_1$ = (1/3 * 1 + 1/9 * 8 + 1/27 * 64) / (1 + 8 + 64) = 0.0492. This first mean can split all the 73 squares into two unbalanced parts: a small portion of the large squares (nine squares larger than the mean) in the head, and a big portion of the small squares (64 squares smaller than the mean) in the tail. For the nine squares in the head, their average size is calculated by $m_2$ = (1/3 *1 + 1/9 * 8) / (1 + 8) = 0.1358. This second mean can split all nine squares into two unbalanced parts: a small portion of the large squares (one square larger than the mean) in the head, and a large portion of the small squares (eight squares smaller than the mean) in the tail (Table 1). The above calculation indicates that the pattern of far more small squares than large ones recurs twice, and therefore ht-index = 2 + 1 = 3. The recurring scaling pattern is also shown in the nested rank-size plots in Figure 1. The ht-index is indeed 3 because there are only three scales: 1/3, 1/9, and 1/27. If we added 512 squares of the smaller size 1/81, the ht-index would increase by one, but the fractal dimension would remain unchanged. From this, we see how ht-index complements fractal dimension in capturing the complexity of fractals.

As the above example shows, head/tail breaks is quite simple and straightforward, i.e., given that there are far more small things than large ones, split things into a few large and many small, and recursively continue the splitting for the large until the notion of far more small things than large ones is violated. Importantly, head/tail breaks leads to a relaxed definition of fractals: a set or pattern is fractal if the notion of far more small things than large ones recurs multiple times, ht-index >= 3. This new definition based on the head/tail breaks is pretty intuitive, and may help refine our eyes or improve our intuitions for fractals. As remarked by Mandelbrot (1982), the most important instrument of thought is the eye rather than mathematical formula. With the new definition, anyone with little mathematical knowledge can easily rely on his/her intuitions to determine whether something is fractal. Now let us examine whether our intuitions have been improved with reference to the Mandelbrot set in Figure 2.

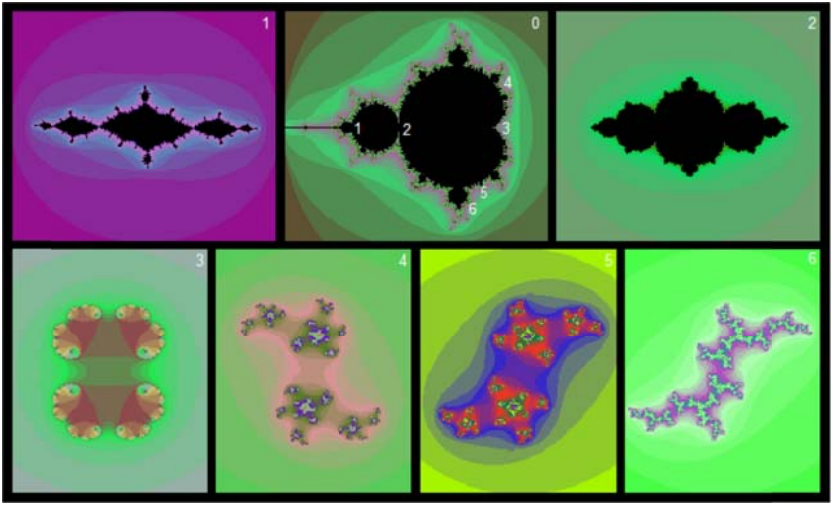

Figure 2: (Color online) Ubiquity of far more small things than large ones in both the Mandelbrot set (Panel 0) and the Julia sets (Panels 1 to 6)
(Note: There are an infinite number of bulbs tangent to the main cardioid of the Mandelbrot set. The Julia sets in Panels 1 and 2 are generated from within the bulbs (black in Panel 0), whereas the Julia sets in Panels 3 to 6 are generated from outside of the bulbs (color in Panel 0). The figures in Panel 0 indicate the approximate locations where the Julia sets are generated.)

It is well known that the Mandelbrot set, probably the most complex shape known to man, comes from the amazingly simple equation: $z = z^2 + c$. Despite the simplicity, we have decided not to consider the underlying mathematics; interested readers can refer to the literature for more details (e.g., Mandelbrot 2004). Instead, we rely on our intuitions by concentrating on the Mandelbrot set shape and an infinite number of convoluted Julia sets shapes it generated, some of which are shown in Figure 2. What the stunning images of these shapes have in common is the ubiquity of far more small things than large ones. The Mandelbrot set can be zoomed into deeply to find similar patterns again and again infinitely, so the Mandelbrot set can be said to be "big data". The Mandelbrot set (Panel 0) contains far more small bulbs than large ones, as do the two related Julia sets (Panel 1-2) generated from within the bulbs (black in Panel 0) of the Mandelbrot set. The Julia sets generated from outside the bulbs (color in Panel 0) of the Mandelbrot set have some dramatically different shapes and colorful images (Panel 3-6), which clearly evoke a sense or intuition that there are far more small structures than large ones. The images also look beautiful. Note that it is essentially not the colors but the underlying fine structures (or recurring pattern of far more small things than large ones) that make the patterns beautiful (Alexander 2002); see Section 4 for further discussion.

### 3. Visualization of city structure and dynamics
When the social scientist Jacob L. Moreno (1934) first studied such human relationships as likes and dislikes, his dream was to map them for a whole city or nation. It now appears that what he dreamed of has been fully realized, not only for a whole nation, but for the entire world, with millions or billions of people connected through social media such as Facebook and Twitter. The New York Times praised Moreno's work as a new human geography (Jones 1933), because the map metaphor was used for portraying the acquired human relationships, with nodes for individuals, links for relationships between the individuals, red lines for liking, black lines for disliking, triangles for boys, and circles for girls. This semiology, together with the methods of data collection and data analysis, were typical social science methods in the age of data scarcity, or the so-called small data era. Nowadays, we have entered the big data era, in which we are overwhelmed by crowdsourcing data, accumulated in social media and contributed by individuals (Goodchild 2007, Kwak et al. 2010, Gao and Liu 2014). In addition, advanced geospatial technologies have already produced a large amount of geographic information such as satellite images (National Research Council 2003). Big data requires new ways of thinking in order to better understand the underlying social and geographic structure and how the structure evolves over time (Jiang 2015). In this connection, visualization offers a powerful means to reach the better understanding.

**3.1 Natural cities derived from POI**
Points of interest (POI) are spread across countries, particularly within cities, represent interesting locations or facilities such as churches, schools, shops, and pubs. As a wiki-like collaboration to create a free editable map of the world, OpenStreetMap (Bennett 2010) has integrated millions of POI, including basic categories such as automotive, eating and drinking, government and public services, health care, and leisure. In this study, I took approximately 2 million POIs for the three European countries: France, Germany, and the United Kingdom from CloudMade (http://download.cloudmade.com/). Following the same procedure of extracting natural cities introduced in the previous work (Jiang and Miao 2014), we built a huge TIN for each country, and then derived natural cities for further scaling analysis. Table 2 presents the basic statistics about the derived natural cities. France, for example, has 280,117 POI, of which 254,008 unique points were used to generate a huge TIN with 835,009 edges. There are far more short edges than long edges, so the distribution is clearly L-shaped. I applied the head/tail division rule (Jiang and Liu 2012) into the massive number of edges, which resulted in two unbalanced parts: those above the mean in the head, and those below the mean in the tail. All those edges in the tail were aggregated, leading to the 9,391 natural cities. Figure 3 shows the resulting natural cities (Panel 1), together with those for the other two countries (Panels 2 and 3). Germany is the densest country in terms of both POI and natural cities, followed by the UK.

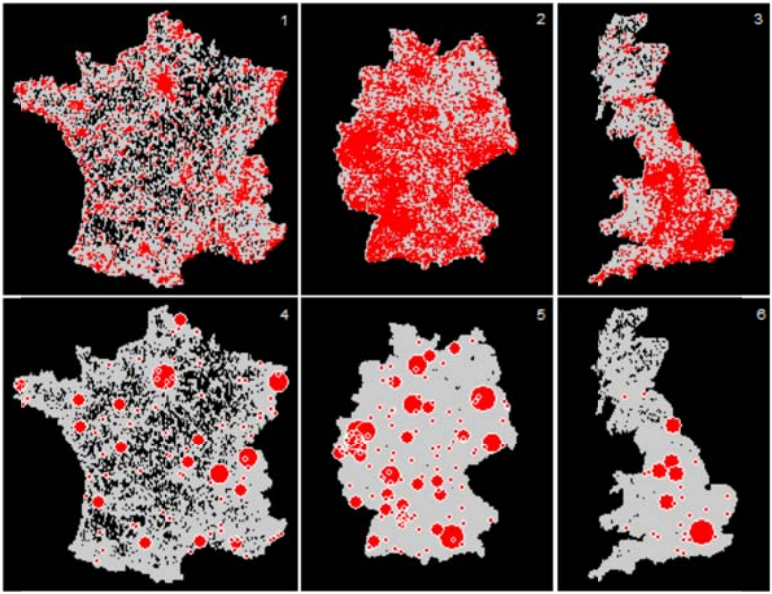

Figure 3: (Color online) The natural cities derived from POI of France (Panels 1 and 4), Germany (Panels 2 and 5), and the UK (Panels 3 and 6)
(Note: The red patches indicate the natural cities or their boundaries, whereas the red dots indicate classified city sizes in terms of the number of POI. As mentioned in this paper, only a few top classes based on the head/tail breaks are shown for visual clarity. The gray background is the points of interest. The map scales are 1:15M.)

Table 2: Basic statistics about the natural cities derived from POI

|  | France | Germany | UK |
|---|---|---|---|
| POI | 280,117 | 1,299,638 | 505,051 |
| Unique POI | 254,008 | 977,357 | 462,424 |
| TINEdge | 835,009 | 3,238,695 | 1,511,023 |
| Natural cities | 9,391 | 48,830 | 16,814 |
| Ht-index/hierarchy | 6 | 7 | 6 |
| Hierarchy shown | 4 | 4 | 3 |

There are far more small natural cities than large ones in terms of the numbers of POI they contain. To effectively visualize the underlying scaling hierarchy of the natural cities, I applied the head/tail breaks to computing the ht-index that is shown in Table 2. France, Germany, and the UK, respectively, have six, seven, and six hierarchical levels or classes. If all the classes were displayed by different sizes of red dots, no matter how small they are, the patterns would not be recognizable. Instead, I chose the top four or three classes (Panels 4, 5, and 6 of Figure 3), which reflect the same scaling patterns of all the classes in the sense that the pattern of far more small things than large ones is retained. The fact that the top classes reflect the whole is the true power of head/tail breaks. In other words, the top classes retain the same scaling pattern of far more small things than large ones of all the classes.

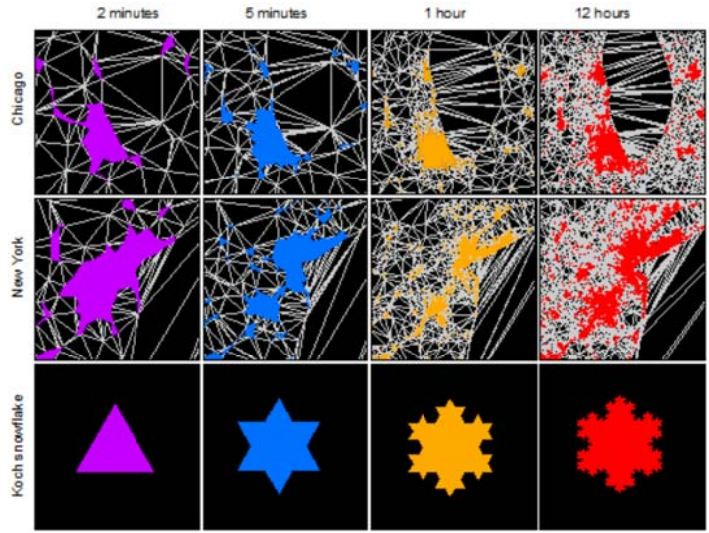

Figure 4: (Color online) The evolution of the natural cities on the background of TIN versus iteration of Koch flake
(Note: A few large pieces become more fragmented, whereas many small pieces are continuously added. Eventually there are far more small cities than large ones. This way of evolution looks very much like that of Koch flake. The major difference between the natural cities and Koch flake is the former being statistically self-similar, and the latter strictly self-similar. The map scales are 1:8.4M.)

### 3.2 Natural cities from tweets locations

Like POI, social media users' locations can be aggregated to form individual natural cities. Unlike POI, Twitter users' geolocations contain very precise time information, up to minutes or seconds. In this way, we can slice the tweets location data minute by minute, hour by hour, in order to track how the natural cities evolve. The derivation of the natural cities followed the same procedure in the previous work (Jiang and Miao 2014), and was based on the fact that there are far more low-density than high-density areas. That is, we generated a huge TIN for the unique locations of tweets and then split the TIN edges into two unbalanced parts: those above the average in the head, and those below the average in the tail. Eventually, those edges in the tail are aggregated into the thousands of natural cities. The procedure is a simple application of the head/tail breaks, or that of the head/tail division rule. Let us consider the four snapshots to examine the underlying fractal structure and nonlinear dynamics of the natural cities (Figure 4). The evolution of the natural cities shows little difference from that of the Koch flake: the former being statistically self-similar, and the latter being strictly self-similar. Accordingly, I claimed that social media could act as a good proxy for studying the evolution of real cities, in order to understand how they are generated and evolve through local and global interactions from the bottom up. This insight could fundamentally change the ways we studied cities in the small data era of the past.

The scaling patterns appear at different levels of geographic space. This is the true sense of ubiquity of fractal geographic features. It appears at a country level, a regional level, and a city level. Figure 5 presents an illustration of the ubiquity of scaling patterns. The large number of natural cities derived from tweets locations are classified into six classes. I display only the top 4 classes for visual clarity, yet they reflect the pattern of the whole set (Panel 0). The enlarged regions of Chicago and New York (Panels 1 and 2) clearly show that there are far more small cities than large ones. At the city level, I compute the connectivity of individual streets, and the degree of connectivity clearly shows a heavy-tailed distribution. All the streets are therefore classified and visualized based on the head/tail breaks. I invite the reader to compare Figures 2 and 5: the former being purely mathematical, and the latter geographic; the former being infinite, and the latter finite; the country as a whole equivalent to the Mandelbrot set as a whole, whereas the cities as parts equivalent to the Julia sets as parts; a city as a whole equivalent to the Mandelbrot set a whole, whereas the streets as parts equivalent to the Julia sets as parts. From the comparison, we see a nested or cascading structure for both the Mandelbrot set and the geographic space, and importantly the shared recurring scaling pattern of far more small things than large ones.

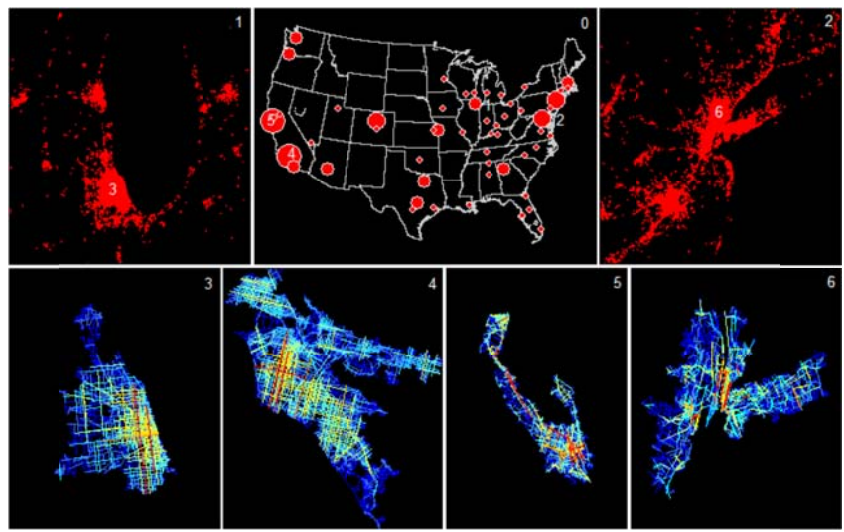

Figure 5: (Color online) Ubiquity of scaling patterns at different levels using the USA (mainland) as an example
(Note: The largest 55 natural cities in the top four classes at the country level (Panel 0); there are far more small cities than large ones at the regional level (Panels 1 and 2), and far more less-connected streets than well-connected ones at the city level (Panels 3 to 6), with blue being the least connected and red the most connected. The map scales for the USA (Panel 0), the regions (Panels 1, and 2), and the cities (Panels 3-6) are respectively 1:60M, 1:6M, and 1:2M.)

### 3.3 Natural cities from nighttime images: China urbanization during 1992-2012
Natural cities can also be automatically extracted from the nighttime images by merging or grouping all those pixels whose values are higher than a certain mean averaged from all pixels. It is important to note that the nighttime images represent stable lights emitted from cities or human settlements in general, and the unstable lights such as fires and other ephemeral nights have been removed already (Elvidge et al. 2014). For a nighttime image of the world or a large country, there are certainly far more dark/gray pixels than light ones. The light pixels (lighter than an average) constitute individual natural cities, for example, about 30,000 natural cities derived for the entire the world (Jiang et al. 2014). To track China (mainland) urbanization in the past two decades or so, I have extracted large numbers of natural cities from the 21 nighttime images each in a year during 1992 and 2012 (Figure 6). The number of natural cities during the period increases dramatically from 518 to 3006 respectively with years 1993 and 2012. As sees from Panel 5, years 1992 and 2010 are two tipping points. The China urbanization in terms of the city numbers is obviously a nonlinear process, by which I mean the process at the time dimension is not steadily growing, but with some bursts, something like a stock price. Instead of showing all 3006 cities with five classes (which is impossible), the China map (Panel 0) shows only top three classes that present clearly a sense of feeling of far more small cities than large ones. This is very much like the USA map in Figure 5 earlier. The regional maps (Panels 1-4) demonstrate the burst behavior of the urbanization near Shanghai, Guangzhou, Beijing, and Fuzhou. The fragmented pieces and irregular city boundaries indicate the fractal nature of the natural cities.

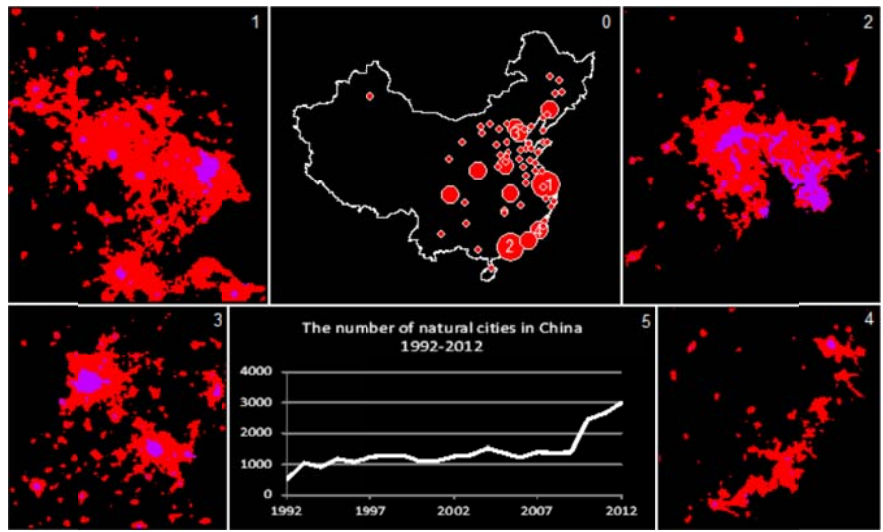

Figure 6: (Color online) China (mainland) urbanization during 1992-2012
(Note: The 3006 natural cities in year 2012 are classified into five hierarchical levels based on head/tail breaks, but the map in Panel 0 shows only three top levels. The natural cities in the four largest and rapidly developing regions: Shanghai (Panel 1), Guangzhou (Panel 2), Beijing (Panel 3) and Fuzhou (Panel 4) are shown in purple for 1992 and in red for 2012. The plot in Panel 5 shows how the number of natural cities increases dramatically during the two decades. The map scales for China (Panel 0), and the regions (Panels 1, 2, 3 and 4) are respectively 1:80M, and 1:6M.)

Let us zoom into some local regions of China to assess how the urbanization has unfolded in space and over time. The five largest and rapidly developing regions (Shanghai, Guangzhou, Beijing, Fuzhou, and Shantou) in the seven time instants are shown in Figure 7 for a close look. It is clear that (1) the city boundaries look irregular, and (2) there are far more small cities than large ones. Given one of the two facts, the natural cities are fractal. This fractal nature of the natural cities can also be seen from the time dimension. There are in total 518 cities in year 1992, five of which are shown in the first column of Figure 7, and the rest of which are far too small to be shown, indicating far more small cities than large ones. This scaling or fractal nature is valid for all other columns of Figure 7. It is also clear that (1) the Shanghai region exceeds that of Guangzhou (including Hong Kong and Macau) and Beijing around 2010, becoming number one, and (2) both the Shanghai and Beijing regions have growth much faster than that of Guangzhou.

Unlike the conventional cities whose boundaries are hard to delineate and measure, the natural cities are accurately defined and delineated in terms of the very definition of natural cities. This is one of the advantages of working with big data. The reader may argue that it is hard to acquire the populations in the natural cities. This is indeed true, but populations are roughly estimated in essence rather than accurately measured like the physical extents of natural cities. Given the estimated nature of populations, I believe that the physical extents of the natural cities are a better indicator than population for urbanization and other related studies. It is impossible for a big country like China to have a census every year, but it is completely possible to obtain as many as possible nighttime images a year. Nighttime images, or the natural cities in particular, provides a better means to track the physical extents rather than populations of human settlements.

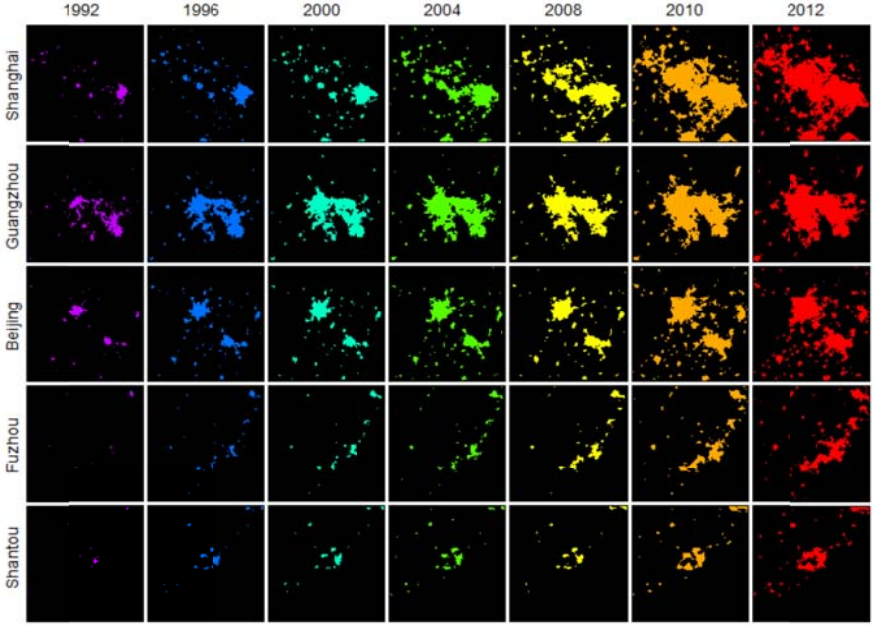

Figure 7: (Color online) The seven snapshots of the five largest and rapidly developing regions during 1992 and 2010
(Note: The natural cities were derived from nighttime images during the period for the five largest and rapidly developing regions near Shanghai, Guangzhou, Beijing, Fuzhou, and Shantou. The snapshots or patterns look fractal in space, and nonlinear in time. It appears that the growth of the Shanghai is faster than that of Guangzhou, although the Guangzhou region is bigger than the Shanghai region in 1992. In addition, Beijing and Tianjin are getting together, so are Fuzhou and Xiamen. The map scales are 1:12.5M.)

**4. Further discussions on head/tail breaks**
Head/tail breaks offers a new, less strict way of looking at our surrounding phenomena, in particular societal and organization phenomena. A phenomenon or structure is fractal if there are far more small things than large ones in it. The notion of far more small things than large ones is not just in terms of geometric properties, but for topological and semantic properties as well, i.e., far more unpopular things than popular things, or far more meaningless things than meaningful things. Consequently, many societal and organizational phenomena are fractal, because they tend to be divided in an unbalanced way, which is known as the 80/20 principle (Koch 1998). Head/tail breaks, in particular the nested rank-size plots, provides a new interpretation of self-similarity. Conventionally, self-similarity refers to the property that the whole has the same shape as one or more of its parts (Mandelbrot 1982). Now the self-similarity can be interpreted by the repeated presence of far more small things than large ones, or alternatively, the repeated appearance of the small head and long tail division. It is the self-similarity that makes visualization of city structure and dynamics possible.

Head/tail breaks applies to data with a heavy-tailed distribution. The heavy-tailed distribution includes power laws as well as lognormal and exponential distributions. Strictly speaking, the exponential distribution is excluded from the heavy-tailed distribution, because its tail is quite short. I included it in the heavy-tailed distribution family because for most real-world data, there is a minimum threshold above which data are claimed to be power laws, lognormal or exponential distributions (Newman 2005). However, while conducting the head/tail breaks for the data, we consider all the data values including those below the minimum threshold. Thus, the data with an exponential distribution including those below the minimum would be heavy-tailed, and can therefore be head/tail broken multiple times. It is also widely recognized that the bigger the data, the more likely they are heavy-tailed.

Fractal structure, or the recurring scaling pattern of far more small things than large ones, possesses a new kind of beauty that positively impacts human well-being (Jiang and Sui 2014). The new kind of beauty, initially discovered and defined by Christopher Alexander (2002), differs fundamentally from conventional wisdom about aesthetics, being personal and subjective. The fractal beauty exists in deep structure, being objective and universal in nature. In other words, it is not the surface colors but the deep fractal structure that makes fractals beautiful. This deep structure is a kind of order that exists not only in nature but also in what we build and make (Alexander 2002), not only in science but also in humanities and social sciences. The beauty, the order revealed by head/tail breaks, or fractal geometry in general, cuts across multiple sciences and disciplines, bridging the two cultures (Snow 1959) to form the third culture. I believe that the visualization examples of city structure and dynamics shown in the paper embody some spirits of the third culture.

Many natural and societal phenomena demonstrate fractal structure and nonlinear dynamics (Mandelbrot and Hudson 2004). To better understand the complexity of social structure and dynamics, we must rely on a range of complexity modeling tools such as fractal geometry, chaos theory, and agent-based simulations (Miller and Page 2007) rather than conventional linear methods such as Euclidean geometry and Gaussian statistics. We must harness the large amounts of data accumulated on social media and the Internet for mining individual and collective behaviors. As a timely response to the challenges arising from big data, the emerging field of computational social science (Lazer et al. 2009, Watts 2007) has been fundamentally transforming the conventional social sciences into both data- and computational-intensive science. Unlike computational sciences in the twenty century, computational social science is a product of the twenty-first century, and it should be correctly interpreted as data-intensive computational social science in the big data era. This is the same for computational geography, which appeared first in the 1990s (Openshaw 1998), should be characterized as computational- and data-intensive in the twenty-first century. In the big data era, cartography faces the same challenge of how to efficiently and effectively visualize the large amounts of crowdsourcing geographic information, and nighttime images. I believe that recognition of the fractal nature of maps and mapping (Jiang 2014) offers a way to meet the challenge.

**5. Conclusion**
This paper has developed an argument that head/tail breaks can be used for visualization of structure and dynamics of natural cities in the big data era. To support the argument, I have developed several case studies applied to natural cities derived from crowdsourcing data and time-series nighttime images. This paper has also discussed how head/tail breaks leads to a new definition of fractals, helping improve our intuitions for seeing fractals in nature and society. Throughout the paper, we have seen both mathematical fractals (such as the Sierpinski carpet, Mandelbrot set, and Koch flakes) and geographic features (such as the natural cities and streets) share the same recurring scaling of far more small things than large ones. The scaling property is what drives the development of head/tail breaks. It is the scaling property that makes head/tail breaks an efficient and effective visualization tool for revealing city structure and dynamics. The power of head/tail breaks lies in its simplicity: split things around an average into a few large and many small, respectively in the head and the tail of the nested rank-size plots, and recursively continue the splitting process in the head until the condition of far more small things than large ones is violated. The simple head/tail breaks and its induced ht-index can help even the general public to see a variety of fractals in science, art, and society.

The notion of natural cities, as a product of the big data era, provides a powerful tool to study human activities on the earth's surface, and enables us to obtain new insights into geographic information harvested from crowdsourcing data, and nighttime images. Compared with conventional real cities that are imposed by authorities from the top down, natural cities are defined from the bottom up, and from individual people and their interactions. Unlike real cities, natural cities can be naturally and objectively derived and delineated from big data such as VGI, social media data, and nighttime images. This makes natural cities universally available for the entire world, i.e. all the natural cities in the world rather than those in some countries. In this regard, big data is probably not so much about bigness, but rather completeness. Natural cities may fundamentally change the ways cities were studied.

**Acknowledgement**
The initial idea of this paper grew out of the talks delivered at Technion's Faculty of Architecture and Town Planning, and University College London's SpaceTimeLab. XXXXX.